\documentstyle[twoside,fleqn,espcrc2,fps]{article}

\newcommand{\AmS}{{\protect\the\textfont2
  A\kern-.1667em\lower.5ex\hbox{M}\kern-.125emS}}

\title{Truncated eigenvalue equation and long wavelength behavior of
  lattice gauge theory
}

\author{Shuo-Hong Guo, Qi-Zhou Chen, Xiyang Fang, Jinming Liu, \address{CCAST
        (World Laboratory),  P.O. Box 8730, Beijing 100080, China,\\
                                     and Department of Physics, Zhongshan
                                        University, Guangzhou 510275, China
                                        }
Xiang-Qian Luo,\address{HLRZ, Forschungszentrum,
D-52425 J{\"u}lich, Germany,\\
and Deutsches Elektronen-Synchrotron DESY, D-22603 Hamburg, Germany}
and Weihong Zheng \address{School of Physics, University of New South Wales,
Kensington New South Wales 2033, Australia
}
                                      }

\begin{document}

\begin{abstract}
  We review our new method, which might be the most direct and efficient
  way for approaching the continuum physics from Hamiltonian lattice gauge
  theory. It consists of solving the eigenvalue equation with a truncation
  scheme preserving the continuum limit. The efficiency has been confirmed
  by the observations of the scaling behaviors for the long wavelength
  vacuum wave functions and mass gaps in (2+1)-dimensional models and
  (1+1)-dimensional $\sigma$ model even at very low truncation orders.
  Most of these results show rapid convergence to the available Monte Carlo
  data, ensuring the reliability of our method.
 \end{abstract}

\maketitle

\section{INTRODUCTION}

The main purpose of our work is to approach the scaling region and extract
physical results by analytic calculations.
For a lattice calculation, a basic requirement is that for weak enough
coupling,
the dimensionless quantities should satisfy the scaling law,
predicted by renormalization group equation.
For $\rm{SU(N_c)}$ gauge theories in 3 dimensions, superrenormalizabilty
and
dimensional analysis tell us that the dimensionless masses $aM$
should scale as
\begin{eqnarray}
{aM \over g^2} \to {M \over e^2}.
\label{s2p1}
\end{eqnarray}
For
(2+1)-dimensional compact \rm{U(1)} and (3+1)-dimensional non-abelian gauge
theories, $aM$ should scale exponentially as
\begin{eqnarray}
{aM} \to exp(-b/g^2).
\label{s3p1}
\end{eqnarray}
If the calculated $M$ data converge
to a stable value, we can get an estimate for the mass.

There have been various analytic methods available in the literature (for
a review see \cite{Guo}).
The main difficulty of the conventional   methods (e.g. strong coupling
expansion) is that
they converge very slowly and very higher order $1/g^2$ calculations are
required
to extend the results to the intermediate coupling region.
Unfortunately,  high order calculations are difficult in practice.

Recently, we proposed a new method \cite{GCL,CGZF,QCD3,MASS} for
Hamiltonian lattice gauge theory. This method consists of
solving the eigenvalue equation with a suitable truncation scheme preserving
the continuum limit.
Even at low order truncation,
clear scaling
windows
for the physical quantities in most cases
have been established, and the results
are in perfect
agreement with the available Monte Carlo data.
Here we review only the work on $\rm{U(1)}_3$, $\rm{SU(2)}_3$ and
2 dimensional
$\sigma$ model, while that for \rm{SU(3)} has been summarized in \cite{Luo}.

\section{THE METHOD}

The Schr{\"o}dinger
equation $H \vert \Omega \rangle = \epsilon_{\Omega} \vert \Omega \rangle$
on the Hamiltonian lattice
for the ground state
\begin{eqnarray}
  \vert \Omega \rangle = exp \lbrack R(U) \rbrack \vert 0 \rangle
\label{b1}
\end{eqnarray}
and vacuum energy $\epsilon_{\Omega}$ can be reformulated as
\begin{eqnarray*}
\sum_{l} \lbrace [E_l,[E_l,R(U)]]+[E_l,R(U)][E_l,R(U)] \rbrace
\end{eqnarray*}
\begin{eqnarray}
- {2 \over g^4} \sum_{p} tr(U_p+U_{p}^{\dagger})
={2a \over g^2} \epsilon_{\Omega}.
\label{schr}
\end{eqnarray}
To solve this equation, let us write $R(U)$
in order of graphs $G_{n,i}$, i.e.,
$  R(U)=\sum_{n} R_{n}(U)=\sum_{n,i} C_{n,i} G_{n,i}(U)$.
Substituting it to (\ref{schr}), we have the $N$th order truncated
eigenvalue equation
\begin{eqnarray*}
\sum_{l} \lbrace [E_l,[E_l,\sum_{n}^{N} R_{n}(U)]]
\end{eqnarray*}
\begin{eqnarray*}
+\sum_{n_1+n_2 \le N}[E_l,R_{n_1}(U)][E_l,R_{n_2}(U)] \rbrace
\end{eqnarray*}
\begin{eqnarray}
- {2 \over g^4} \sum_{p} tr(U_p+U_{p}^{\dagger})
={2a \over g^2} \epsilon_{\Omega}.
\label{b2}
\end{eqnarray}
By taking the coefficients of the graphs $G_{n,i}$ in this equation to zero,
we obtain a set of non-linear algebra equations, from which
$C_{n,i}$ are determined.
The similar method applies to the eigenvalue
equation for the mass and its wave function \cite{MASS}.
Therefore, solving lattice field theory is reduced to solving the
algebra equations.

The lowest order graph is quite simple: $R_1(U)=C_{1,1}(U_p +h.c.)$
The first term in (\ref{b2}) doesn't generate new graphs, but the
second term
does, i.e.
\begin{eqnarray*}
  [E_l,G_{n_1}(U)] \in R_{n}(U) + lower ~ orders,
\end{eqnarray*}
\begin{eqnarray*}
  [E_l,G_{n_1}(U)][E_l,G_{n_2}(U)] \in R_{n_1+n_2}(U)
\end{eqnarray*}
\begin{eqnarray}
  +lower ~ orders.
\end{eqnarray}

Two questions arise:

\noindent
1) Should all the new graphs generated by the second term in (\ref{b2})
be taken as independent
graphs of order $n_1+n_2$? For abelian gauge theories, the answer is yes.
For non-abelian gauge theories, because of the uni-modular conditions
\cite{QCD3,MASS,Luo,GCFC}, there is a mixing problem
not only  for the graphs of the same order, but also for graphs of
different orders.
The classification for independent graphs is particularly
complecate for \rm{SU(3)}.

\noindent
2) For $n_1+n_2 > N$, should we keep the lower order graphs in
$[E_l,G_{n_1}(U)][E_l,G_{n_2}(U)]$?
To preserve the correct limit,
at $Nth$ order truncation,
one should to DROP all these graphs.
This is the essential feature of our truncation scheme,
which differs sufficiently from the scheme in \cite{Green}.

There have also been some other truncation schemes proposed in
\cite{SW,Bishop}.
One of their major problems is the violation of the
long wavelength structure or continuum limit of the equation, and
consequently
the violation of the scaling law (\ref{s2p1}) or (\ref{s3p1})
for the physical quantities.

Let's see further why the equation (\ref{b2}) should be truncated in the way
suggested
 at point 2).
The continuum limit of a graph $G_{n,i}(U)$ is
\begin{eqnarray*}
  G_{n,i}(U)=e^2 a^4[A_{n,i} ~ tr ({\cal F}^2)
  \label{small_a}
\end{eqnarray*}
\begin{eqnarray}
  +a^2 B _{n,i} ~ tr ({\cal D} {\cal F})^2+...]
\end{eqnarray}
with
${\cal F}$ the field strength tensor and ${\cal D}$ the covariant derivative.
It has been generally proven \cite{GCL} that in the continuum limit
the second term of (\ref{b2}) term should behave as
\begin{eqnarray}
  [E_l,G_{n_1}(U)][E_l,G_{n_2}(U)]
\propto e^2 a^6 ~Tr({\cal D} {\cal F}_{\mu,\nu})^2.
\end{eqnarray}
To preserve this correct limit, when the equation
(\ref{b2}) is truncated to the $Nth$ order, all the graphs created by
$[E_l,R_{n_1}(U)][E_l,R_{n_2}(U)]$ for $n_1+n_2 \le N$ must be considered.
On the other hand, all the graphs created by this term for $n_1+n_2 > N$
should be dropped, even there are lower order graphs.
Otherwise the partial sum
of the lower order graphs would make this term behave
in a considerably different
(wrong) way.

\section{RESULTS}

Once the coefficients $C_{n,i}$ are obtained by solving (\ref{b2}),
we can use
(\ref{b1}) and (\ref{small_a})  to compute the parameters $\mu_0$ and $\mu_2$
in the vacuum wave function
for the long wavelength
configurations $U$ \cite{Arisue}
\begin{eqnarray*}
\vert \Omega \rangle=exp \lbrack - {\mu_0} \int d^{D-1}x ~ tr {\cal F}^2
\end{eqnarray*}
\begin{eqnarray}
- {\mu_2} \int d^{D-1}x  ~tr ({\cal D} {\cal F})^2 \rbrack.
\label{a1}
\end{eqnarray}

\begin{figure}[htb]
\fpsxsize=7.5cm
\vspace{-20mm}
\fpsbox[70 90 579 760]{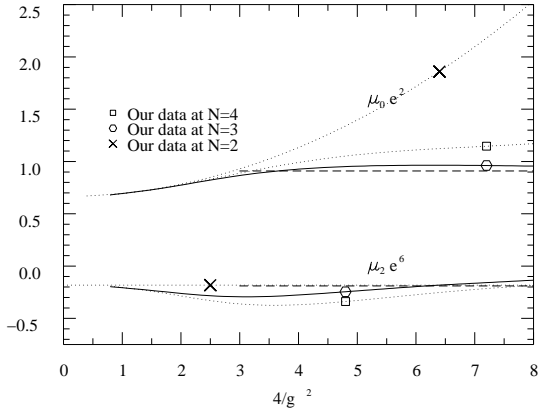}
\vspace{-10mm}
\caption{Parameters in the vacuum wave function of $\rm{SU(2)}_3$ model.
The dashed lines show the mean values for the Monte Carlo data.}
\label{fig1}
\end{figure}

The results for  $\mu_0$ and $\mu_2$ in $3$ dimensional \rm{SU(2)}
gauge theory are shown in Fig. 1. Empressively, nice scaling behavior is
obtained even at $N=3$,
and the data for $4/g^2 > 4$ are in good agreement with the Monte Carlo
measurements \cite{Arisue}. The order $N=4$ data
are also included in this figure.
Although there are no big differences between
the results at $N=3$ and $N=4$,
higher order
calculation seems necessary to ensure that the results would finally
occur at the correct values.

Figure 2 shows our results for the mass gap in $\rm{SU(2)}_3$ at different
truncation orders. For comparison, the results from the truncation method
of Llewellyn Smith and Watson
(LS-W) \cite{SW} and $14th$ order series expansion \cite{Zheng} are
also included.
Again, even at $N=3$, with satisfaction of the scaling law,
our results are in best agreement with the data taken from
the continuum limit of the Monte Carlo data \cite{Teper}.

\begin{figure}[htb]
\fpsxsize=7.5cm
\vspace{-20mm}
\fpsbox[70 90 579 760]{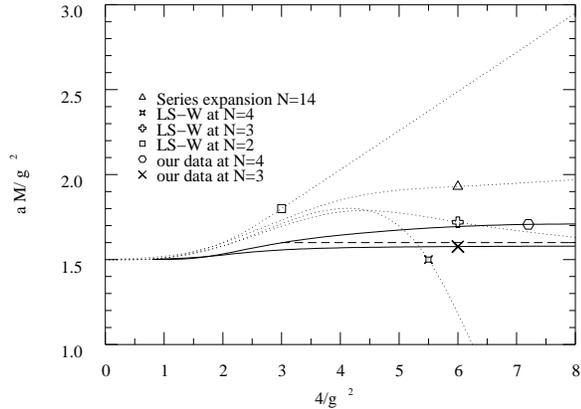}
\vspace{-10mm}
\caption{The mass gap of the $\rm{SU(2)}_3$ model
from three different methods.
  The dashed lines show the continuum limit of
  the Monte Carlo data.}
\label{fig2}
\end{figure}

As is mentioned above, for non-abelian gauge theories, the uni-modular
conditions lead to the existence of different choices for
independent graphs. In \rm{SU(2)}, because of $trU_p^{\dagger}=trU_p$,
all the disconnected graphs can be transformed into the connected ones,
used as an independent set of the graphs.
Our results in Figs. 1 and 2 are from such a choice, while the comparison
between different
choices (connected, disconnected \cite{GCL} and inverse)
has been made in \cite{GCFC}. Of course, one is free
to choose arbitrary set
of independent graphs. A criterion for a good choice is that
it is convergent more rapidly
to the continuum limit than other ones at lower order truncation.
This has been obviously demonstrated for $\rm{SU(3)}$ \cite{Luo,ASSYM}.

The most intriguing scaling law is the exponential scaling (\ref{s3p1}).
Before investigating \rm{QCD} in 3+1 dimensions, we would like to test our
method in a (2+1)-dimensional compact \rm{U(1)} model, which has many
properties of
the realistic theory.
Here there is no ambiguity induced by uni-modular
conditions in $\rm{SU(N_c)}$
  theories.
  Because the nature of the abelian group greatly
simplifies the calculations,
  we can easily write a program at arbitary orders.

\begin{figure}[htb]
\fpsxsize=7.5cm
\vspace{-20mm}
\fpsbox[70 90 579 760]{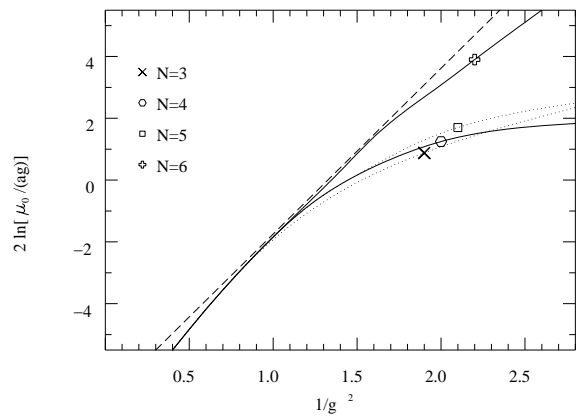}
\vspace{-10mm}
\caption{Relevant quantity  for  $\mu_0$ in the long
  wavelength
  vacuum state of compact
  $\rm{U(1)}_3$. The dashed line is the expected scaling law.}
\label{fig3}
\end{figure}

\begin{figure}[htb]
\fpsxsize=7.5cm
\vspace{-20mm}
\fpsbox[70 90 579 760]{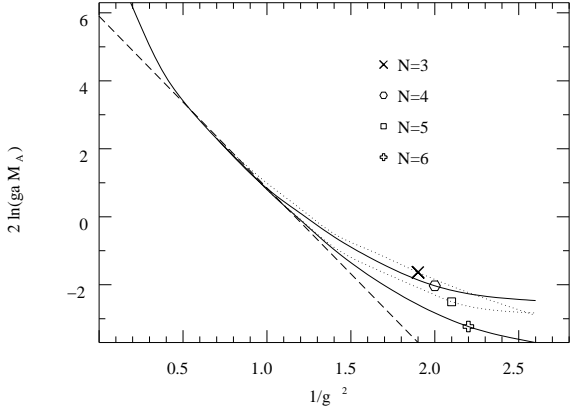}
\vspace{-10mm}
\caption{Relevant quantity for the asymmetric mass $M_A$
  of compact
  $\rm{U(1)}_3$. The dashed line is the expected scaling law.}
\label{fig4}
\end{figure}

The results for $\rm{U(1)}_3$ are shown in Figs. 3 and 4 respectively.
At relatively low
truncation orders (comparing to 16th order series expansion \cite{Zheng}),
a scaling window has been seen around $1/g^2=1$,
and such a window becomes wider with the order $N$.
Most impressively, there is an obvious tendency to converge to
the scaling curve (dash line). This implies that
the obtained physical quantities
occur at their correct values.
Fitting the N=6 data in the scaling region, we get \cite{U1}

\begin{eqnarray*}
   {\mu_{0} \over ag} =3.1120 \times 10^{-2}  exp(2.54(3)/g^2),
\end{eqnarray*}
\begin{eqnarray}
   (M_{A} a g)^2 =365(73) exp( -5.0(2)/g^2),
 \end{eqnarray}
or $\mu_0 M_A=0.59(5)$.

The non-linear (1+1)-dimensional $\sigma$ model is
another interesting application of our method.
According to the theoretical expectation,

\begin{eqnarray}
   M a \propto {1 \over g^2} exp(-{2 \pi \over g^2}  -{g^2 \over 8 \pi}).
 \end{eqnarray}
 Calculations at $N=5,6,7,8$ have been carried out.
 Preliminary results are quite encouraging, and they are in reasonable
 agreement with the Monte Carlo data \cite{Ha}.
To reach the asymptotic scaling
 region, higher order calculation seems necessary.

\section{SUMMARY}
The reason for the success of our method  is that
the truncated eigenvalue equation
preserves the continuum limit. The results for
(2+1)-dimensional models and (1+1)-dimensional $\sigma$ models are presented
to support the efficiency and reliability of our method.
In conclusion, the eigenvalue equation with a proper truncation scheme
may be the most direct and efficient way for
extracting the continuum physics.

The members at Zhongshan are supported by Inst. High
Education, and XQL is sponsored by DESY. We thank the discussions
with H. Arisue, P. Cai, C. Hamer, D. Sch{\"u}tte and A. Sequ{\'\i}.

\end{document}